\documentclass[sigconf]{acmart}
\usepackage{booktabs} 
\usepackage{enumitem}
\usepackage{amsmath}
\usepackage{algorithm}
\usepackage[noend]{algpseudocode}
\usepackage{algpseudocode}
\usepackage{soul}
\usepackage{color}
\usepackage{multirow}
\usepackage{subcaption}
\usepackage{graphicx}
\usepackage{lipsum}
\usepackage{soul}
\usepackage{graphicx,subcaption}

\DeclareMathAlphabet{\pazocal}{OMS}{zplm}{m}{n}

\def\BibTeX{{\rm B\kern-.05em{\sc i\kern-.025em b}\kern-.08emT\kern-.1667em\lower.7ex\hbox{E}\kern-.125emX}}

\copyrightyear{2019}
\acmYear{2019}
\setcopyright{none}
\acmConference[KDD '19]{The 25th ACM SIGKDD Conference on Knowledge Discovery and Data Mining}{August 4--8, 2019}{Anchorage, AK, USA} \acmBooktitle{The 25th ACM SIGKDD Conference on Knowledge Discovery and Data Mining (KDD '19), August 4--8, 2019, Anchorage, AK, USA}
\acmPrice{15.00} \acmDOI{10.1145/3292500.3330781} \acmISBN{978-1-4503-6201-6/19/08}

\begin{document}

\title{Topic-Enhanced Memory Networks for Personalised Point-of-Interest Recommendation}

\author{Xiao Zhou}
\affiliation{%
	\institution{Computer Laboratory }
	\institution{University of Cambridge }
	\city{Cambridge}
	\country{UK}
}
\email{xz331@cam.ac.uk}

\author{Cecilia Mascolo}
\affiliation{%
	\institution{Computer Laboratory }
	\institution{University of Cambridge }
	\city{Cambridge}
	\country{UK}}
\email{cm542@cam.ac.uk}

\author{Zhongxiang Zhao}
\affiliation{%
	\institution{WeChat Business Group}
	\institution{Tencent Inc. }
	\city{Beijing}
	\country{China}}
\email{abnerzxzhao@tencent.com}

\renewcommand{\shortauthors}{X. Zhou et al.}

\begin{abstract}
Point-of-Interest (POI) recommender systems play a vital role in people's lives by recommending unexplored POIs to users and have drawn extensive attention from both academia and industry. Despite their value, however, they still suffer from the challenges of capturing complicated user preferences and fine-grained user-POI relationship for spatio-temporal sensitive POI recommendation. Existing recommendation algorithms, including both shallow and deep approaches, usually embed the visiting records of a user into a single latent vector to model user preferences: this has limited power of representation and interpretability. In this paper, we propose a novel topic-enhanced memory network (TEMN), a deep architecture to integrate the topic model and memory network capitalising on the strengths of both the global structure of latent patterns and local neighbourhood-based features in a nonlinear fashion. 
We further incorporate a geographical module to exploit user-specific spatial preference and POI-specific spatial influence to enhance recommendations. The proposed unified hybrid model is widely applicable to various POI recommendation scenarios. Extensive experiments on real-world WeChat datasets demonstrate its effectiveness (improvement ratio of 3.25\% and 29.95\% for context-aware and sequential recommendation, respectively). Also, qualitative analysis of the attention weights and topic modeling provides insight into the model's recommendation process and results. 
\end{abstract}



\maketitle

\section{Introduction}

In the information explosion era, recommender systems have become increasingly important in our daily life by improving the suggestions we received online based on personal tastes~\cite{zhou2018micro}.
As a particular type of recommender systems focusing on real-world locations, point-of-interest (POI) recommendation has drawn intensive attention recently with the prevalence of smart mobile devices and the rapid growth of location-based social networks (LBSNs). Specifically, LBSNs, such as Yelp\footnote{www.yelp.com} and Foursquare\footnote{https://foursquare.com/} link the physical and virtual worlds by offering users a way to share their life experiences with POIs via the "check-in" function. Through the lens of check-in data in LBSNs, rich knowledge becomes available to mine users' visiting preferences for customised POI recommendation~\cite{zhang2015geosoca}. Such kind of services not only benefit users by providing them appealing venues to explore, but also facilitate targeted advertising with significant economic efficiency enhancement.


Unlike other recommender systems that push digital goods, e.g. e-books, news, and movies, POI recommendation aims at offering users preferred new venues to explore in the physical world~\cite{wang2018exploiting}, which can be largely affected by various real-life factors and thus faces more challenges. Firstly, to experience a POI, a user has to physically visit it, which is generally more costly and time-consuming than watching a movie or listening to a song online. Also, even if the user does visit the venue, she might prefer not to leave a check-in record due to privacy and security concerns~\cite{xie2016learning}. For these reasons, the number of POIs that a user generally interacts with is extremely small compared with the total number of POIs in the LBSN, leading to a very sparse user-POI matrix that plagues a considerable number of current POI recommender systems \cite{yin2017spatial}. Moreover, unlike pure online interactions, users' activities in the physical world are limited by travel distance and time~\cite{wang2018exploiting}, making POI recommendation a more complex issue. Considering the distinguishing characteristics of POI recommendation task, corresponding research efforts have been spend improving its effectiveness.
The existing approaches in this area can be roughly grouped into two categories: neighbourhood-based methods and latent factor models~\cite{ebesu2018collaborative}. Neighbourhood-based approaches are good at describing local strong associations between users and POIs, but typically ignore the vast majority of user-POI interactions that are not similar enough. While for latent factor models like Matrix Factorization (MF)\cite{koren2009matrix}, they can easily extract the global structure of relationships between users and POIs, but have limitations in capturing their local features. Instead of establishing either a neighbourhood-based or latent factor model, we propose a unified hybrid architecture which combines the benefits of both techniques to enrich predictive capabilities by capturing both global and neighbourhood features of users and POIs. Furthermore, aiming to learn higher order complex relations between users and POIs that cannot be obtained through simple functions, deep learning techniques are adopted.

The model proposed in this paper, \textit{\textbf{T}opic-\textbf{E}nhanced \textbf{M}emory \textbf{N}etworks} (\textbf{TEMN}), is an end-to-end framework for personalised POI recommendation, which consists of three integrated components. Specifically, a memory-augmented neural network is employed to model fine-grained relations between each user-POI pair via a metric learning approach. This memory network (MN) component is linked with a topic modeling module, called temporal latent Dirichlet allocation (TLDA) through user embedding. In this way, TLDA module can not only be used to infer the general preferences of users towards POIs and time slots, but also influence the representations of POIs and users in MN to enhance the recommendation performance. Furthermore, to capture geographical effects in POI recommendation, geographical modeling module is also devised. 

Our primary contributions can be summarised as follows:
\begin{itemize}[leftmargin=0.3cm]
	\item We propose an end-to-end deep learning framework that integrates neighbourhood-based and global preferences of users. 
	\item We devise a more flexible architecture that incorporates multiple types of contextual information into POI recommendation and make it applicable to various recommendation scenarios. 
	\item We build a hybrid model which combines supervised and unsupervised learning and capitalises on the advances in both memory networks and topic modeling. Through a mutual learning mechanism, our model is also able to provide users' probability distributions over topics that are influenced by memory network. 
	\item Comprehensive experiments on large WeChat datasets in different recommendation scenarios demonstrate the effectiveness of TEMN against competitive state-of-the-art baselines (improvement ratio of 3.25\% and 29.95\% for context-aware and sequential recommendation, respectively).  
	\item Besides quantitative improvements, by incorporating neural attention mechanisms and topic model in TEMN, the interpretability of the POI recommendation is significantly promoted.
\end{itemize}
\section{Related Work} \label{related_work}

To alleviate the sparsity issue of user-POI interaction data and cater for the context awareness nature of POI recommendation, auxiliary information have been incorporated into existing POI recommender systems \cite{wang2018exploiting}. For instance, \cite{lian2014geomf} incorporated spatial clustering features of check-ins into a weighted matrix factorization framework. \cite{xie2016learning,feng2017poi2vec,zhao2017geo} claimed that POIs clustered within the same geographical region should be restricted to similar representations compared with their distant counterparts in recommendation. Another group of researchers focused on how temporal effects can be utilised to improve POI recommendation by considering either temporal cyclic patterns \cite{gao2013exploring,zhang2014lore} or sequential influence \cite{cheng2013you,chen2014nlpmm,zhao2017geo}. Besides spatio-temporal information, other types of information have also been explored to facilitate POI recommendation performance, such as social influence \cite{tang2013exploiting,cheng2012fused,ye2011exploring, zhang2015geosoca,cho2011friendship,li2016point,zhang2016gmove}, categorical information \cite{zhang2015geosoca,zhao2015sar}, visual content \cite{wang2017your}, and text information \cite{chang2018content,wang2017location} of POIs. 

From the angle of recommendation techniques, scholars have proposed some effective models. For example, MF \cite{koren2009matrix} and its variants have been widely applied to POI recommendation tasks. Vanilla MF was initially used to deal with general POI recommendation, where only user-POI interactions are leveraged. When additional information like spatio-temporal information and social relationship became available, some models derived from basic MF (e.g., IRenMF \cite{liu2014exploiting}, GeoMF \cite{lian2014geomf}, GeoIE \cite{wang2018exploiting}) were proposed and adapted to context-aware recommendation. Markov chain models (e.g., LBPR \cite{he2017category}, NLPMM \cite{chen2014nlpmm}, FPMC-LR \cite{cheng2013you}, LORE \cite{zhang2014lore}) and recurrent neural network (RNN) (e.g., ST-RNN \cite{liu2016predicting}, CARA \cite{manotumruksa2018contextual}, DRCF \cite{liu2016predicting}) have also been employed when temporal effects are particularly valued in sequential POI recommendation.
\section{Problem Formulation}\label{Problem_form}
\newcommand{\La}{\pazocal{L}}
\newcommand{\Ua}{\pazocal{U}}
\newcommand{\Va}{\pazocal{V}}
\newcommand{\Oa}{\pazocal{O}}

In this section, we formulate our research problems and introduce the notations used throughout the paper. 

In a POI recommender system, let $\Ua$ and $\Va$ denote the user set and the POI set, respectively. The user-POI interaction matrix $Y = \left \{ y_{uv}\mid u\in \Ua, v\in \Va \right \}$ is defined according to users' implicit feedback, where each entry $y_{uv}$ in $Y\in \mathbb{R}^{|\Ua|\times |\Va|} $ records whether user $u$ has visited POI $v$ as follows:

\begin{equation}
y_{uv}=
\begin{cases}
1,& \text{if interaction $(u,v)$ is observed;}\\
0,& \text{otherwise.}
\end{cases}
\end{equation}

\noindent Here a value of 1 for $y_{uv}$ indicates that user $u$ has visited the venue $v$ before. Otherwise, it is assigned a value of 0, meaning that there is no interaction existing between them and the preference of user $u$ for POI $v$ is currently unclear. Mathematically, the main goal of the POI recommender system is to estimate the scores of the unobserved entries in matrix $Y$ for recommendation.

Since our aim is to propose a more generic framework for POI recommendation that is sufficiently flexible to incorporate various contextual information and is widely applicable to different scenarios, we present the definitions of three POI recommendation scenarios involved in this paper as follows:  

\noindent\textbf{\textit{General POI recommendation.}} In this scenario, only the interactions between users $\Ua$ and POIs $\Va$ are taken into account. In other words, the interaction matrix $Y$ is taken as the only input to the recommender system. The main task here is to provide each $u\in \Ua$ a list of POIs consisting of venues that the target user has not explored before, but is potentially interested in.  

\noindent\textbf{\textit{Sequential POI recommendation.}} Sequential recommendation predicts successive POIs that are likely to be visited by a user given her check-in history. In such a scenario that sequence matters, for each user $u$, her past interactions with POIs are firstly sorted according to the check-in timestamps ascendingly. Instead of using the whole visiting sequence of the user directly, we divide it into some shorter segments by setting a time interval threshold $\Delta T $. Specifically, if the time interval between two adjacent check-ins is larger than $\Delta T $, they are not regarded as successive check-ins, since the large time interval may indicate they are irrelevant \cite{zhang2014lore}. Based on this assumption, a check-in sequence set forms for each user and can be used as training data for sequential POI recommendation.

\noindent\textbf{\textit{Context-aware POI recommendation.}} As stated in Section \ref{related_work}, multiple types of rich contextual information associated with check-ins can be used to enhance POI recommender systems. In this paper, we will focus more on spatio-temporal aware POI recommendation. 
\section{Topic-enhanced Memory Networks}\label{Proposed_model}
In this section, our proposed model Topic-Enhanced Memory Networks for POI recommendation (TEMN) is introduced. After presenting the general architecture of the model, we will describe its main components respectively before elaborating on how to integrate them with each other for overall training.   

\subsection{General Framework}
\begin{figure*}[h]
	\centering
	\includegraphics[width=17cm]{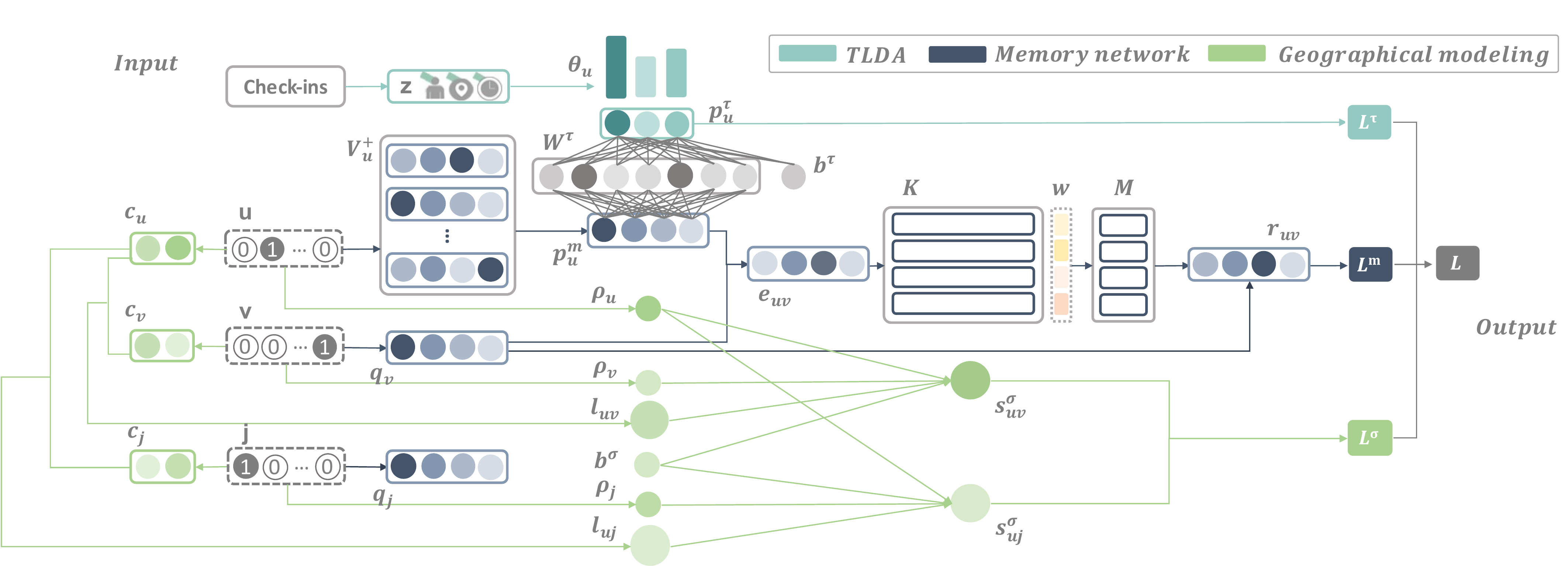}
	\caption{The overall architecture of the TEMN. It consists of three modules represented by different colours. A user, a visited POI, a negative sampling POI, and the user's visiting history are taken as input for POI recommendation.}
	\label{fig:architecture}
\end{figure*}

In Figure \ref{fig:architecture}, we present a visual depiction of TEMN's overall architecture. It can be seen that at a high level, TEMN consists of three key parts: memory network, TLDA, and geographical modeling. The first two components are linked with each other, which allows for modeling the nonlinear interactions between local features learned from the neighbourhood-based memory network and global preferences extracted from the topic model. With such an architecture, given a user $u$, the final prediction score for each unvisited POI $v$ is a reflection of both the fine-grained relationship between user-POI pair <$u,v$> and the general interest of user $u$. Technically, this capability is realised by the topic-enhanced user embedding, which is made up of two parts: 1) a representation related to the user's memory component that encodes her visiting records (named as memory embedding), and 2) a vector employed to encode her intrinsic preference extracted through topic modeling (named as intrinsic embedding).

\subsection{Memory Network}\label{MN}
In our framework, the module of memory network is utilised to encode users' previous check-in records, capture the local relationships between venues, and learn the relational vector for each user-POI pair which represents complex interactions between them. 

\subsubsection{User and POI embedding}
In the memory network, POIs are initially represented by binary sparse vectors through one-hot encoding. Then they are further transformed into low-dimensional dense vectors to get POI embeddings. Formally, given a venue $v$, let $q_v\in\mathbb{R}^d$ denotes its embedding vector. Here $d$ is the dimension size of the POI embedding. Different from POI embedding method, for each user $u$, her memory embedding $p_u^m\in\mathbb{R}^d$ is derived based on the POIs she has visited $V_u^+$ as:

\begin{equation}
p_{u}^{m}=\frac{1}{|\Va_{u}^{+}|}\sum q_{i}^{m}  (i \in  \Va_{u}^{+})
\end{equation}

\noindent where $q_{i}^{m}$ is the embedding vector of POI $i$. Then for each given user-POI pair <$u,v$>, we calculate their joint embedding $e_{uv}$ through Hadamard product as below:

\begin{equation}
e_{uv} = p_u^m\odot q_v
\end{equation}

\noindent The dimension size of the generated vector $ e_{uv}$ is the same as that of $p_u^m$ and $q_v$. This joint embedding of user-POI pair will be fed into a memory-augmented neural network later to make predictions. 

\subsubsection{Key-value memory networks}

The core of the memory network module consists in a memory matrix $\textbf{M}\in \mathbb{R}^{h\times d}$, where $h$ represents the number of memory slots and $d$ denotes the size of each memory cell. In our model, this memory matrix stores the information related to user-POI interactions and can be read and updated adaptively by adopting key-value attention mechanisms. Technically, a key matrix $\textbf{K} \in \mathbb{R}^{d \times h} $ is employed, where the number of key vectors is $h$, and the dimensionality of each key slot $k_i \in \textbf{K}$ is $d$, which is the same as that in user and POI embedding. Our main aim here is to learn an attention vector $w$. For each <$u,v$> pair, when their joint embedding $e_{uv}$ is fed into the networks, the similarity between $e_{uv}$ and each key vector $k_i$ will be first calculated by dot product before being converted to a relevance probability using softmax function:

\begin{equation}
w_i = softmax(e_{uv}^T\cdot k_i ) = \frac{exp(e_{uv}^T\cdot k_i)}{\sum_{j} exp(e_{uv}^T\cdot k_j))},\forall j=1,2,\cdots h
\end{equation}

\noindent where $w_i$ represents the attention weight of the $i$-th element in the attention vector $w$, which will be further utilised to collect useful information from each memory slice $m_i \in\textbf{M}$ to generate the relation vector $r_{uv}$ of the given user-POI pair <$u,v$> through:

\begin{equation}
r_{uv}=\sum_{i} w_{i} m_i ,\forall i=1,2,\cdots h
\end{equation}

This obtained latent relation vector $r_{uv}$ is the output of the memory network module. 

\subsubsection{Optimisation} The latent representation of user-POI interactions plays as a translation vector that enables the model to learn relationships between each specific user-POI pair in metric space adaptively \cite{tay2018latent}. The scoring function implemented here is based on the idea of TransE \cite{bordes2013translating} model, where we compute the loss of the triple <$ p_u^m, r_{uv}, q_v$> by the Euclidean distance and calculate the score for each user-POI pair <$u,v$> as:

\begin{equation}
s_{uv}^m=-||p_{u}^{m} + r_{uv} - q_{v}||_2^2
\label{equ:Score_uvm}
\end{equation}

\noindent where $\left \| \cdot  \right \|_2$ is the $L2$-norm of the vector. $s_{uv}^m$ represents the score of user-POI pair <$u,v$> predicted through the memory network. The higher the score, the more likely user $u$ would be interested in the recommended venue $v$. 

We optimise the memory network module under the Bayesian personalised ranking (BPR) \cite{rendle2009bpr} criterion, where the key assumption is that observed entries should be ranked higher than their unobserved counterparts. Given a positive instance $v \in \Va_{u}^{+}$ visited by user $u$, a corrupted example $j \notin  \Va_{u}^{+}$ which the user has not interacted with yet would be selected based on a certain negative sampling rule. Then both of the POIs $v$ and $j$ would go through the same user and POI embedding layer as introduced, respectively. When calculating the score $s_{uj}^m$ for the corrupted pair <$u,j$>, the relation vector used is the same as that generated by positive pair <$u,v$>. The reason for doing like this is that we empirically found it providing much better performance than generating another relation vector for the negative sample. Based on pairwise comparisons, the objective function in this part is defined as:

\begin{equation}
\La^m = \sum_{u} \sum_{v\in \Va_{u}^{+}} \sum_{j \notin  \Va_{u}^{+}} max(0, s_{uj}^m-s_{uv}^m+\lambda^m )
\end{equation}

\noindent where $\lambda^m $ is the margin that separates positive and negative examples. Rectified linear unit ($ReLU$) function is adopted by us as the nonlinear activation function here as it gives better experimental results than using the sigmoild function. 

\subsection{Temporal Latent Dirichlet Allocation}

Apart from user's memory embedding $p_u^m$ achieved through the memory network, which captures neighbourhood-based interests of a user, we think the intrinsic preferences of a user is also essential in creating a comprehensive user profile for personalized POI recommendation. For this reason, we employ temporal latent Dirichlet allocation (TLDA) \cite{zhou2018discovering} in this section to mine the inner interests of users from a more general perspective. 

The TLDA model is an unsupervised machine learning algorithm that characterises each user as a mixture of patterns with both venue and temporal preferences considered. In the TLDA model, users' check-in data containing timestamp and POI information are taken as the input. Here the set of a user's historical check-ins as a whole is viewed as a document, where POIs and time slots associated with the check-ins are regarded as words analogously. As a generative probabilistic model in nature, TLDA models the underlying generative process whereby the user's check-in set is created to infer her probability distribution over several latent patterns. In Table~\ref{tab:tlda}, the notation and description of the parameters employed in the model to fit observed check-in data are provided. Next we will give a brief explanation about how the intrinsic embeddings of users are estimated during this synthetic process. 

To begin with, the pattern distributions of user $ \theta _{u} $ and time $ \phi_{t} $ are sampled from a prior Dirichlet distribution parameterised by $\alpha$ ($\theta _{u}\sim Dir(\alpha)$) and $ \gamma $ ($\phi_{t} \sim Dir(\gamma)$), respectively. Then the pattern assignment $z_{ut}$ of each POI can be drawn from a multinomial distribution $z_{ut}\sim Multinomial\left (\theta _{u},\phi_{t} \right)$ as below:

\begin{equation}
P(z_{ut}|\alpha ,\gamma )=\int_{\theta _{u}}P(z_{u}|\theta _{u})P(\theta _{u}|\alpha)\int_{\phi_{t}}P(z_{t}|\phi_{t})P(\phi_{t}|\gamma)
\end{equation}

\noindent where $z_u$ is the chosen pattern of user $u$; and $z_t$ is the chosen pattern of time $t$. Then a venue $v_{ut}$ is chosen from a multinomial probability $ \varphi_{z}$ conditioned on the pattern $z$, which is sampled from a prior Dirichlet distribution $ \beta $ ($ \varphi_{z} \sim Dir(\beta)$) by:

\begin{equation}
P(v_{ut}|z_{ut})=\int_{\varphi_{z}}P(v_{ut}|\varphi_{z})P(\varphi_{z}|\beta )
\end{equation}

\noindent Following that, the venue would be generated by estimating the maximum likelihood of $v_{ut}$ by the following equation:

\begin{equation}
P(v_{ut}|\alpha ,\gamma, \beta)=\int_{\theta _{u}}\int_{\phi_{t}}\int_{\varphi_{z}}P(v_{ut},z_{ut},\theta _{u},\phi_{t},\varphi_{z}|\alpha ,\gamma, \beta )
\end{equation}

\noindent Eventually, the probability distribution over $\pi$ patterns for a given user $u$ can be estimated by applying the Gibbs sampling algorithm. It can be seen as a pattern mixture that represents each user's intrinsic embedding denoted as $p_u^\tau$:

\begin{equation}
p_{u }^{\tau}=(P(z_{u}^1|\theta _{u}),P(z_{u}^2|\theta _{u}),\cdots ,P(z_{u}^\pi |\theta _{u}))
\end{equation}

It is worth noting that the TLDA module not only provides us the pattern-user probability distribution, the distributions of venues and time slots associated with each pattern can also be estimated to help us understand what each pattern might refer to.

\begin{table}[h]
	\setlength{\abovecaptionskip}{0 cm}
	\setlength{\belowcaptionskip}{0.35 cm}
	\renewcommand\arraystretch{1.25}
	\small
	\begin{tabular}{||c||c||}
		\hline 
		Notation&Description\\
		\hline
		\hline  
		$ \alpha $& Dirichlet prior over the pattern-user distributions\\
		\hline  
		$\beta$ & Dirichlet prior over the venue-pattern distributions\\
		\hline  
		$\gamma$ &Dirichlet prior over the pattern-time distributions\\
		\hline  
		$\theta _{u}$ & pattern distribution of user $ u $\\
		\hline  
		$ \varphi _{z} $ & venue distribution of pattern $ z $\\
		\hline  
		$ \phi_{t} $ & pattern distribution of time $ t $\\
		\hline  
		$ z_{ut} $ & pattern of venue of user $ u $ at time $ t $\\
		\hline  
		$ v_{ut} $ & venue of user $ u $'s check-in at time $ t $\\
		\hline 
	\end{tabular}
	\captionsetup{margin=0cm}
	\caption{Notation and description of parameters in TLDA.}
	\label{tab:tlda}
\end{table}

\subsection{Fusion of MN and TLDA}

In the last two subsections, the modules of memory network and TLDA have been introduced, respectively. Essentially, they are not two separate parts, but an interrelated whole system that is capable of learning both neighbourhood-based and global characteristics of users and POIs, as well as the fine-grained relationships between specific user-POI pairs. In this subsection, we will introduce how the memory network and TLDA can be fused together, and mutually reinforce each other under the TEMN framework.

The key part connecting the two modules is user embedding. More specifically, we define our problem as a multi-class classification task and use a neural network approach to learn the topic-enhanced user embedding for the memory network. As our main aim is to map the memory embedding $p_{u}^{m}$ of a user $u$ to her intrinsic embedding $p_{u }^{\tau}$ through the neural network architecture, $p_{u}^{m}$ is taken as the input data, and the intrinsic embedding $p_{u }^{\tau}$ of the user is applied as the output where each element of the vector represents a class. Mathematically, the model is defined as:

\begin{equation}
\ O^{\tau }= f (W^{\tau}p_{u}^{m} + b^{\tau })
\label{equ:out}
\end{equation}

\noindent where $W^{\tau}$ and $ b^{\tau }$ denote the weight matrix and bias vector, respectively. $ f(\cdot )$ is the activation function. To get the final predicted probabilities $\hat{p}_{u}^{\tau}$, the output vector $O^\tau$ is converted by the softmax function as:

\begin{equation}
\hat{p}_{u}^{\tau} =softmax( O^{\tau })
\end{equation}

\noindent By introducing softmax activation function in the output layer, the cross-entropy loss is used for optimisation:

\begin{equation}
\La^{\tau}=-\sum _{u}\sum _{i}p_{u }^{\tau}(i)\cdot log(\hat{p}_{u}^{\tau}(i)) , \forall i=1,2,\cdots , \pi 
\end{equation}

\noindent where $i$ represents one of the $\pi $ patterns in the TLDA. The selection of activation function $ f(\cdot )$ in Equation~\ref{equ:out} is not necessarily restricted to a particular function. Many options like sigmoid, hyperbolic tangent, and ReLU can be taken in practice according to specific application domain. Alternatively, one can also choose to apply Equation~\ref{equ:out} without an activation function and convert it to a softmax regression model. Here, we choose the ReLU function as it yields the most favourable result.

\subsection{Geographical Modeling}\label{Geo_model}
Given the TEMN is expected to be competent for context-aware recommendation, it is necessary to incorporate spatio-temporal information into the model. As introduced in the last subsection, temporal effects have been considered in the design of TLDA. In this subsection, we are going to present how geographical influence is taken into account in the TEMN. 

Most existing methods assume that the geographical effect between a pair of user and POI is determined by their physical distance, which fails to capture the asymmetry between the user and POI, the high variation of geographical preferences across users, and the differences of geographical influence among POIs. In this sense, we think that to measure the geographical effect only with physical distance is not enough. Instead, we introduce the concepts of \textit{ geographical preference } and \textit{ geographical influence } which are specifically learned for each user and POI so as to depict the fine-grained geographical effect between them. 

For a user $u$, each POI $v \in \Va^+_u $ visited by her has certain geographic coordinates $c_v = (g_{lat}^v, g_{lon}^v)$ recorded simultaneously when the certain check-in was created. Here $g_{lat}^v$ and $g_{lon}^v$ represent the latitude and longitude of venue $v$, respectively. Through an investigation about her check-in history, the central point $c_u$ of user $u$ where her check-ins clustered around geographically can be determined by:

\begin{equation}
c_{u}=(\frac{1}{|V_{u}^{+}|}\sum_{i} g_{lat}^{i},\frac{1}{|V_{u}^{+}|}\sum _{i}g_{lon}^{i}),\forall i\in |V_{u}^{+}|
\end{equation}

\noindent Then the physical distance $l_{uv}$ between user $u$ and venue $v$ can be calculated through:

\begin{equation}
l_{uv}=||c_{u}-c_{v}||_2
\end{equation}

To model the geographical effect between each user-POI pair <$u,v$>, three factors are introduced in this part: 1) the \textit{geographical preference} $\rho_{u}$ of user $u$, which characterises how sensitive the user is to distance; 2) the \textit{geographical influence} $\rho_{v}$, representing the influential level of the venue reflected on physical distance; and 3) the geographical distance between them $l_{uv}$. Based on this, the geographical score $s_{uv}^{\sigma}$ of user-POI pair <$u,v$> is calculated by:

\begin{equation}
s_{uv}^{\sigma} = \rho_{u} l_{uv} +\rho_{v} l_{uv} +  b_{u}^{\sigma}=(\rho_{u}+\rho_{v}) l_{uv}+b^{\sigma}
\label{equ:Score_uvs}
\end{equation}

\noindent where $b^{\sigma}$ denotes the geographical bias. Similarly as that in memory network (Section~\ref{MN}), the objective function applied here is also based on BPR\cite{rendle2009bpr}. Given a venue $v$ which has been visited by user $u$, and a negative venue $j$ that she never interacted with, the pairwise loss function is defined as:

\begin{equation}
\La^{\sigma}=\sum _{u}\sum_{v\in V_{u}^{+}} \sum_{j \notin V_{u}^{+}} max(0, s_{uj}^\sigma-s_{uv}^\sigma+\lambda^{\sigma} )
\end{equation}

\noindent where $\lambda^{\sigma}$ is the margin separating positive and negative examples.

\subsection{Joint Training}
Our complete recommender system is a hybrid of memory network, TLDA, and geographical modeling, which is a topic-enhanced memory network with spatio-temporal contextual information considered. Until now, we have elaborated each module respectively and introduced how memory network is linked with the TLDA. In this part, we present how the TEMN framework functions as a whole. 

For each pair of user-POI <$u,v$>, we can calculate its score achieved from the memory network $s_{uv}^m$ through Equation~\ref{equ:Score_uvm}, and its geographical score $s_{uv}^ \sigma $ by Equation~\ref{equ:Score_uvs}. Then the overall score $s_{uv}$ can be easily obtained through the following equation: 

\begin{equation}\label{all_score}
s_{uv}=s_{uv}^{m} + \eta s_{uv}^{\sigma }
\end{equation}

\noindent where $\eta$ is a weighting parameter, which can be optimally tuned to study the effects of incorporating geographical modeling for POI recommendation under our framework. The overall objective function for TEMN is a weighted sum of all the contributions of each related individual objective as:

\begin{equation}\label{all_loss}
\La = \La^{m}  +\varsigma \La^{\tau}  +\epsilon\La^{\sigma} + \lambda \left \| \varrho   \right \|_2^2
\end{equation}

\noindent where $\La^{m}$, $\La^{\tau}$, and $\La^{\sigma}$ denote the loss induced by the memory network, TLDA, and geographical modeling respectively. $\varsigma$ and $\epsilon$ are weighting parameters specified to influence to what degree the TLDA and geographical effects should be taken into account during the optimisation. $\varrho$ is the model parameter set; and $\lambda$ is a parameter which controls the importance of the last term, where we regularise all the model parameters to prevent overfitting. In the training phase, mini-batch stochastic gradient descent (SGD) is employed to minimise the objective function.

\section{Experiments}
\label{sec:experiments}
The effectiveness of the proposed model for top-N POI recommendation under various scenarios is evaluated in this section. We begin with experimental setup, and then present and analyse the experimental results both quantitatively and qualitatively. 

\subsection{Experimental Setup}
\subsubsection{Datasets}
We conduct our experiments on the WeChat\footnote{http://www.wechat.com/en} Moments check-in data. This study was approved by Tencent\footnote{https://www.tencent.com/en-us/index.html} and the Computer Science and Technology Department University of Cambridge Ethics Committee. Initially, the whole dataset contains all the POI check-in records created by WeChat users living in Beijing for one year spanning from September 2016 to August 2017. After filtering the data, we obtain an evaluation dataset, termed as \textit{WeChat(GPR)} for general and context-aware recommendation. For sequential POI recommendation, it is further processed as that introduced in Section \ref{Problem_form}. This newly created dataset containing the set of successive check-ins for each user is named as \textit{WeChat(SPR)}. The statistics of the two datasets are summarised in Table~\ref{tab:data}. Apart from the interaction information between users and POIs, the timestamp associated with each check-in and the side information of each POI consisting of coordinates and categories (e.g., restaurant, art museum, sports centre) are also available.

\begin{table}[h]
	\setlength{\abovecaptionskip}{0 cm}
	\setlength{\belowcaptionskip}{0.35 cm}
	\renewcommand\arraystretch{1.25}
	\small
	\begin{tabular}{||c||c|c|c|c||}
		\hline 
		Dataset&\#Users&\#POIs&\#Chek-ins&Density\\
		\hline
		\hline  
		WeChat (GPR)&75,973 &28,183 &564,4965 &0.264\% \\
		\hline 
		WeChat (SPR)&28,566 &13,826 & 509,589 & 0.129\% \\
		\hline 
	\end{tabular}
	\captionsetup{margin=0cm}
	\caption{Statistics of the evaluation datasets.}
	\label{tab:data}
\end{table}

\subsubsection{Evaluation Metrics.}
Two common ranking evaluation metrics, \textit{Hit Ratio (HR)} and \textit{Normalised Discounted Cumulative Gain (NDCG)} are adopted. HR measures whether the test POI shows within the top $N$ in the ranked list, and the NDCG takes the position of the test POI into account and penalises the score if it is ranked lower in the list.

\subsubsection{Baselines} We compare our proposed approach against competitive baselines representing latent factor models, topic models, metric learning methods, Markov chain models, and deep learning-based models under various recommendation scenarios.

To evaluate the performance of our model in general POI recommendation, the following methods are considered:
\begin{itemize}[leftmargin=0.5cm]
	\item \textbf{MF} \cite{koren2009matrix} Matrix factorization is one of the most popular model-based collaborative filtering approaches for recommendation, which models the user-POI relationship using inner product. 
	\item \textbf{BPR} \cite{rendle2009bpr} Bayesian personalised ranking optimises the MF model with a pairwise ranking loss. It is tailored to recommendations with implicit feedback data.
	\item \textbf{LDA}\cite{blei2003latent} It is an unsupervised learning approach initially applied to cluster documents and discover topics based on their contents. It is the prototype of our TLDA module that does not consider temporal factors in human mobility.
	\item \textbf{CML} \cite{hsieh2017collaborative} This is a method used to exam whether the latent relational learning of $r_{uv}$ is necessary in our design of memory network module. CML minimises the Euclidean distance between user and POI vectors as $||p_{u}^{m} - q_{v}||_2^2$.
	\item \textbf{LRML} \cite{tay2018latent} This is a memory network-based approach which learns a translation vector between a pair of user and item. It can be seen as a memory network module without topic-enhanced effects and geographical influence. 
	\item \textbf{TEMN(GPR)} It is a variant of TEMN, which reserves the memory network component, replaces the TLDA with LDA, and removes geographical modeling part. 
\end{itemize}
The second group of baselines are models devised for sequential POI recommendation, which include:
\begin{itemize}[leftmargin=0.5cm]
	\item \textbf{LORE} \cite{zhang2014lore} It considers sequential influence and geographical effects in recommendation by adopting additive Markov chain.
	\item \textbf{ST-RNN} \cite{liu2016predicting} This is a RNN-based state-of-the-art POI recommendation model, which considers information about the time interval and distance between pair of check-ins. 
	\item \textbf{TEMN(SPR)} Our complete model with the WeChat (SPR) data fed into, which is employed to test our model's effectiveness in successive POI recommendation.
\end{itemize}
The last group of baseline approaches supporting the integration of contextual information associated with POI check-ins are:
\begin{itemize}[leftmargin=0.5cm]
	\item \textbf{GeoMF} \cite{lian2014geomf} This method extends the MF by augmenting latent factors with the user's activity region and POI's influence area.
	\item \textbf{TLDA} \cite{zhou2018discovering} TLDA is an extension model of LDA that can be applied to extract users' lifestyle patterns with temporal preferences considered. It is also employed as the topic module in our framework to enhance memory network. 
	\item \textbf{TEMN(CPR)} It is our complete model with the WeChat (GPR) dataset employed. 
\end{itemize}

\subsection{Experimental Results}
The comparison results of our proposed model and the baselines on two WeChat datasets in three POI recommendation scenarios are reported in Table~\ref{tab:exp_results}. The key observations from the experimental results are summarised as follows:

\begin{table*}[]
	\setlength{\abovecaptionskip}{-0.35 cm}
	\setlength{\belowcaptionskip}{0.35 cm}
	\renewcommand\arraystretch{1.2}
	\begin{tabular}{||c|c||c|c||c|c||}
		\hline 
		Type&Method&HR@5&NDCG@5&HR@10&NDCG@10\\
		\hline 
		\hline 
		\multirow{6}*{General}&MF&0.50961&0.30345& 0.63174&0.34028\\
		\cline{2-6}
		&BPR&0.52561&0.33747&0.64313&0.37597\\
		\cline{2-6}
		&LDA&0.69221&0.53960&0.80958&0.57778\\
		\cline{2-6}
		&CML&0.67983&0.45613&0.78813&0.49163\\
		\cline{2-6}
		&LRML& $0.69594^*$&$0.54669^*$&$0.81199^*$&$0.58436^*$\\
		\cline{2-6}
		&\textbf{TEMN(GPR)}&\textbf{0.70389}(+1.141\%)&\textbf{0.55221}(+1.010\%)&\textbf{0.81752}(+0.682\%)&\textbf{0.58914}(+0.817\%)\\
		\hline 
		\hline
		\multirow{3}*{Context-aware}&GeoMF&0.63714& 0.50435&0.70742& 0.52741\\
		\cline{2-6}
		&TLDA& $ 0.71518^*$&$ 0.55852^*$ &$ 0.83033^*$ &$0.59601^*$\\
		\cline{2-6}
		&ST-RNN&0.60240&0.47044&0.74372&0.51621\\
		\cline{2-6}
		&\textbf{TEMN(CPR)}&\textbf{0.72876}(+1.899\%)&\textbf{0.57666}(+3.248\%)&\textbf{0.83398}(+0.440\%)&\textbf{0.61053}(+2.436\%)\\
		\hline 
		\hline 
		\multirow{3}*{Sequential}&LORE&$0.53187^*$&0.37880&$0.69148^*$&0.43049\\
		\cline{2-6}
		&ST-RNN&0.53041&$0.42867^*$&0.62798&$0.45999^*$\\
		\cline{2-6}
		&\textbf{TEMN(SPR)}&\textbf{0.62105}(+16.767\%)&\textbf{0.54847}(+27.947\%)&\textbf{0.69769}(+0.897\%)&\textbf{0.57304}(+24.577\%)\\
		\hline 
	\end{tabular}
	\captionsetup{margin=0cm}
	\caption{Performance comparison of different methods in three recommendation scenarios. Best performance is in boldface. We use "*" to mark the best performance from baselines for each comparison and report the improvement ratio of our model over the best baseline performance for each scenario in parentheses.}
	\label{tab:exp_results}
\end{table*}

\begin{figure*}
	\vspace{0.8cm}
	\setlength{\abovecaptionskip}{0 cm}
	\setlength{\belowcaptionskip}{0.2 cm}
	\begin{subfigure}[b]{.246\linewidth}
		\includegraphics[width=\linewidth]{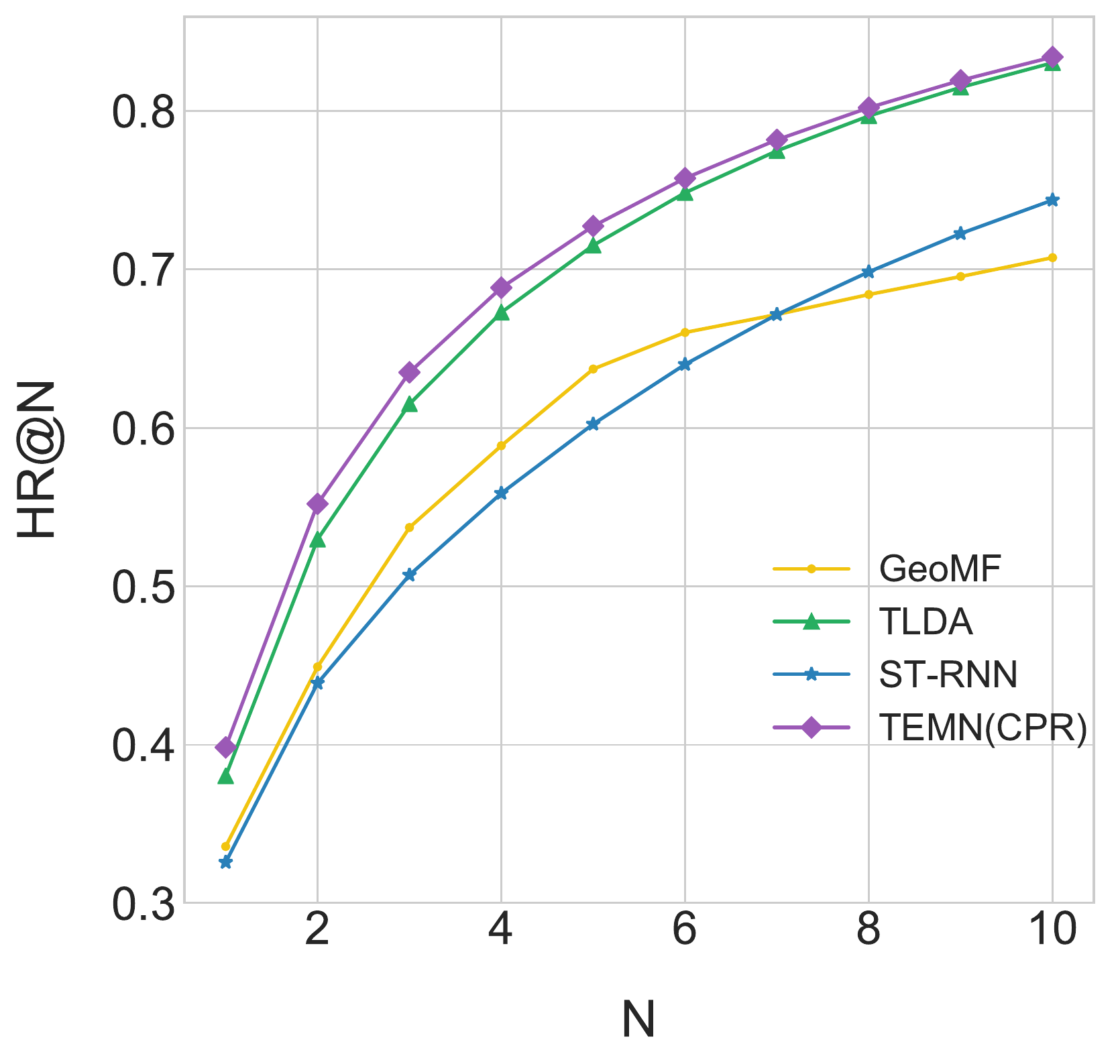}
		\caption{CPR--HR@N}\label{fig:CPR_HR}
	\end{subfigure}
	\begin{subfigure}[b]{.248\linewidth}
		\includegraphics[width=\linewidth]{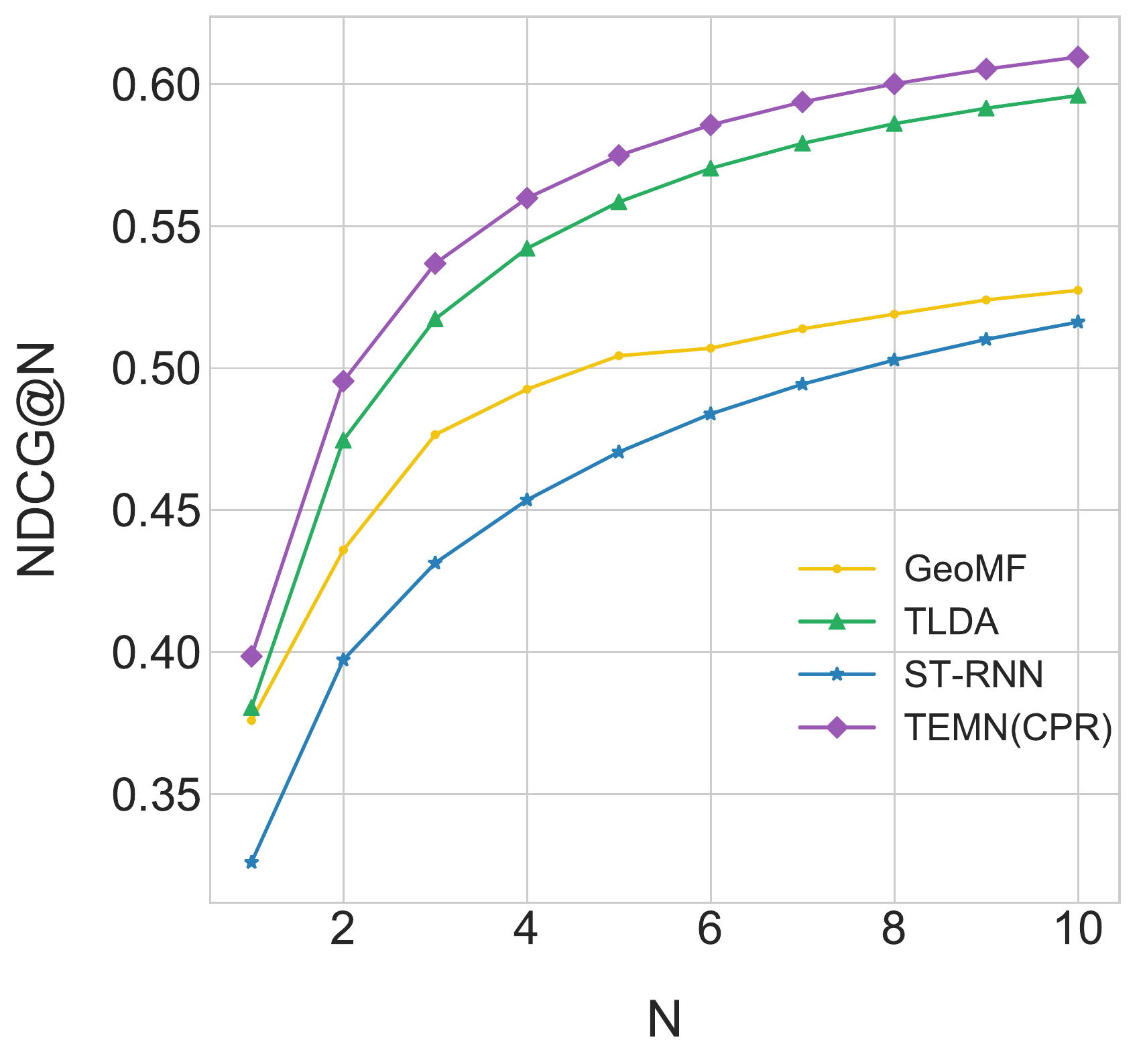}
		\caption{CPR--NDCG@N}\label{fig:CPR_NDCG}
	\end{subfigure}
	\begin{subfigure}[b]{.246\linewidth}
		\includegraphics[width=\linewidth]{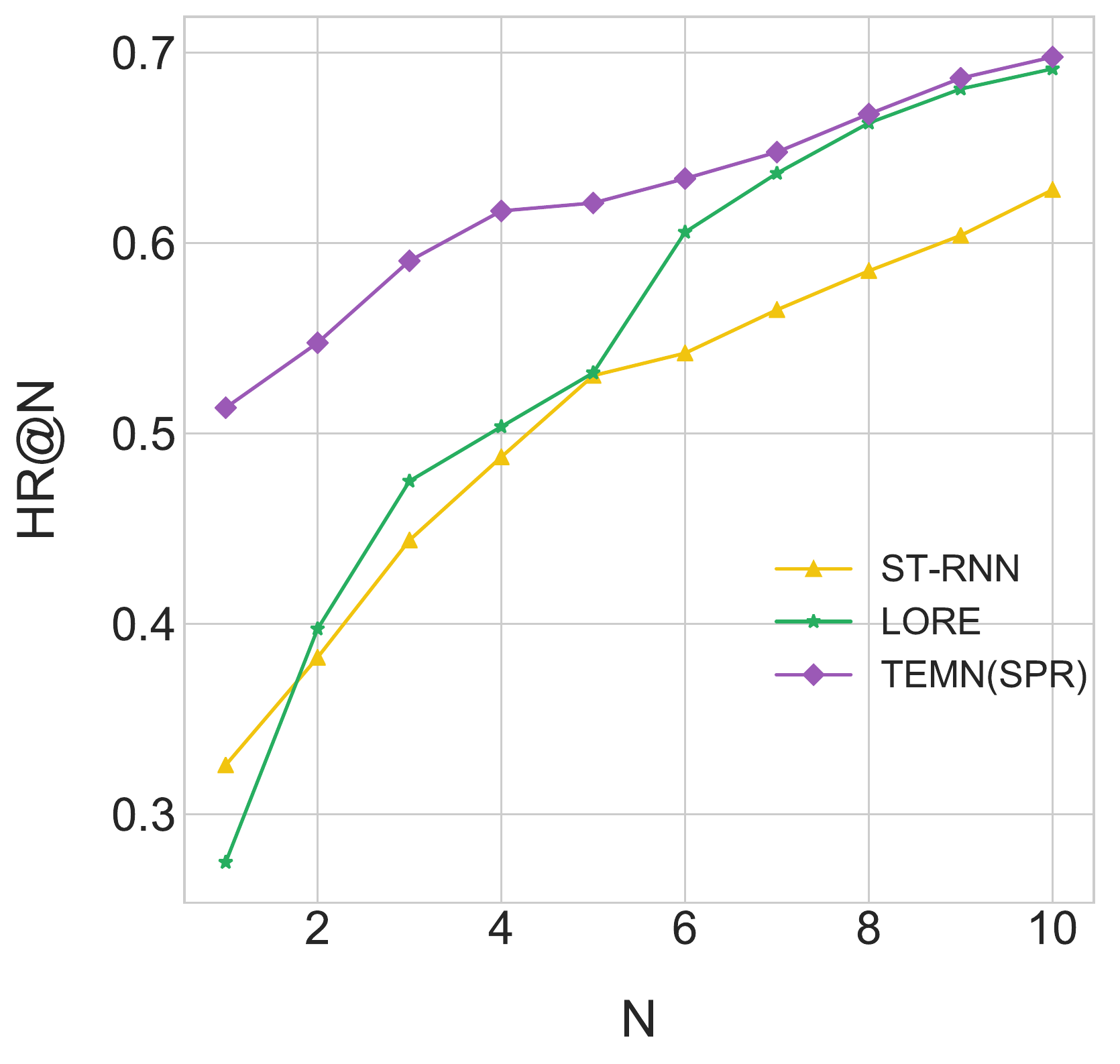}
		\caption{SPR--HR@N}\label{fig:SPR_HR}
	\end{subfigure}
	\begin{subfigure}[b]{.248\linewidth}
		\includegraphics[width=\linewidth]{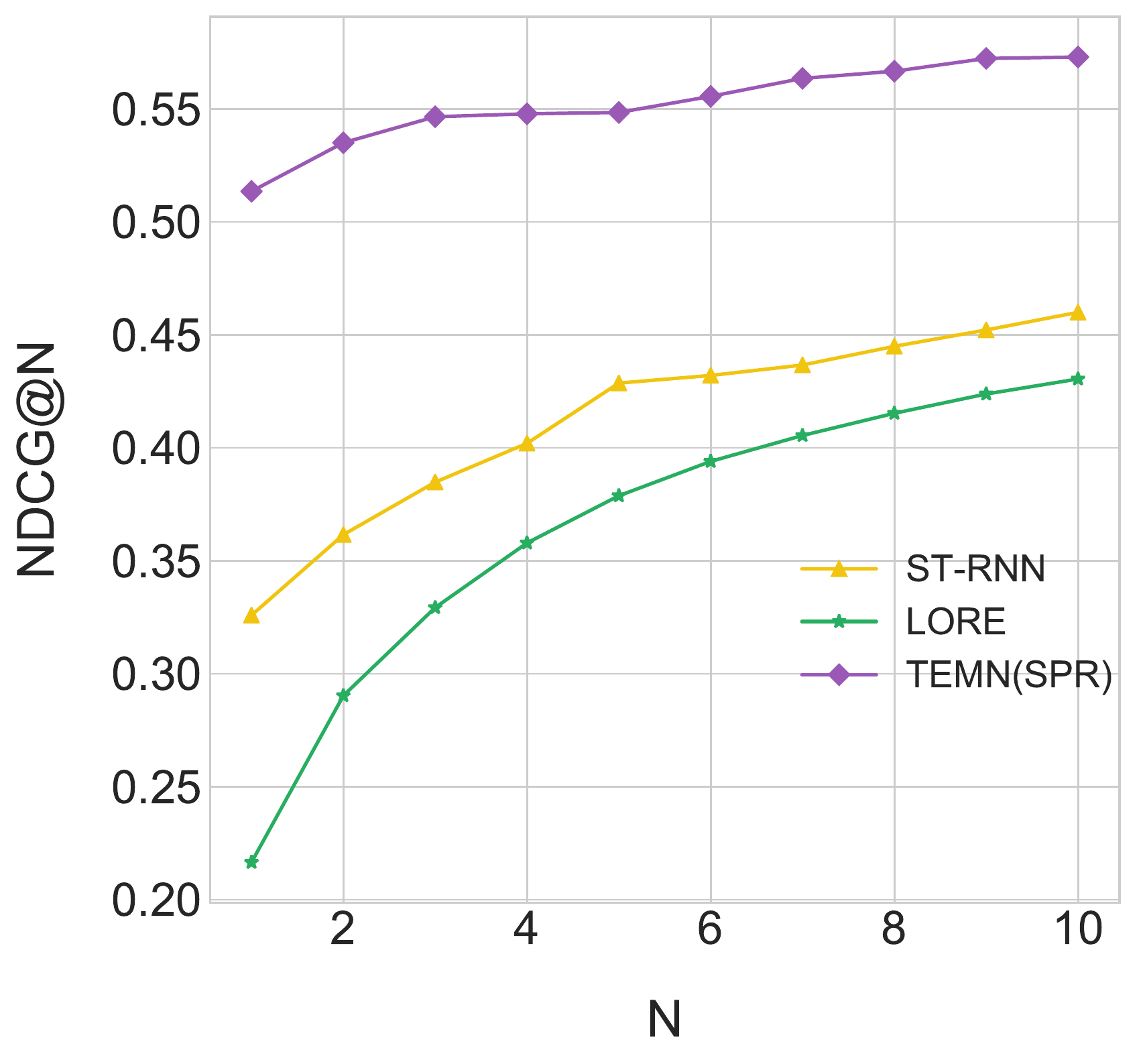}
		\caption{SPR--NDCG@N}\label{fig:SPR_NDCG}
	\end{subfigure}
	\caption{Evaluation of Top-$N$ POI recommendation where $N$ ranges from 1 to 10 on the two WeChat datasets in two scenarios.}
	\label{fig:TOP_N}
\end{figure*}

(1) Encouragingly, it is clear that the performance of our proposed model TEMN is consistently better than all the baselines under different conditions by a relative large margin (the relative improvement ratio over the best baseline in the general, context-aware, and sequential recommendation scenario is 5.16\%, 3.25\%, and 27.95\%, respectively). Additionally, as can be seen from both Table \ref{tab:exp_results} and Figure \ref{fig:TOP_N}, TEMN offers significantly larger improvement for NDCG than HR, suggesting that the proposed model gives a higher ranking to the positive test POI, and shows the feasibility and effectiveness of applying the topic-enhanced memory network to the top-N POI recommendation. 

(2) Considering the three variants of the TEMN, we find that the overall performance order is as follows: TEMN(GPR) > TLDA> LRML. This indicates that an integrated structure (without contextual information) of memory network and topic model performs better than the individual components. To further compare the performance of memory network module (LRML) and topic model module (TLDA), TLDA shows its superiority by giving around 2-3\% improvement across all the metrics and cut offs. TLDA is a model proposed by us last year to discover individuals' lifestyle patterns. This is the first time it is applied to the recommendation task and provides stunning performance by beating all the other baselines. In a nutshell, even though the performance of the individual modules is marginally worse than the complete TEMN model, they provide the best baseline performance in general and context-aware recommendation scenario, respectively. Moreover, the outstanding performance of TLDA highlights the importance of users' long-term interests mining and temporal effect consideration in POI recommendation.

(3) Taking other baseline methods for the first two recommendation scenarios into account, neural network-based approaches (CML, LRML, and ST-RNN) and topic models (LDA and TLDA) outperform MF-based methods (MF, BPR, and GeoMF) in most cases, which indicates the usefulness of capturing more complex user-POI relationships through nonlinear methods and the global preference of users through topic modeling. The relatively poor performance of MF-based approaches suggests that adopting dot product may not be enough to depict user-POI interactions as discussed in \cite{he2017neural}. Another interesting finding is that LRML performs better than CML, proving the advantage of employing a translation vector to capture relationships between a user-POI pair. In addition, GeoMF, TLDA, and TEMN (CPR) all give better recommendation results than their prototypes (MF, LDA, and TEMN(GPR)), indicating the value of incorporating spatio-temporal information into POI recommender systems. The unfavourable performance of ST-RNN may be due to RNN-based models' problem of dealing with long sequences \cite{huang2018improving}. It is a more suitable approach for successive POI prediction, where recent visits are particularly important. 

(4) Utilising a different dataset (WeChat (SPR)), the experimental results of models in the third group for sequential POI recommendation is not directly comparable with that in the other two groups and is discussed individually here. As it can be seen from the table, our proposed model TEMN(SPR) gives the best performance again with dramatic improvement ratios in most cases. These results indicate the effectiveness of TEMN for sequential recommendation, which is actually not surprising because the memory mechanism as well as the geographical modeling design in our architecture provide better expressive power to model recent user-POI interactions and fine-grained geographical factors for each specific user-POI pair. Among the baseline methods applied to sequential POI recommendation task, it is interesting to see that Markov chain model (LORE) performs better for HR, while RNN-based method (ST-RNN) shows better results for NDCG. This finding can be utilised in practical application that when the length of recommendation list is limited ($N$ is small in the top-N recommendation), our TEMN model and ST-RNN are better choices for successive POI recommendation. 

\subsection{Detailed Analysis of the Proposed Model}
After discussing the experimental results quantitatively, in this subsection we present some qualitative observations and analysis results of TEMN.

\subsubsection{Attention Visualisation}

\begin{figure}[h]
	\centering
	\includegraphics[width=6cm]{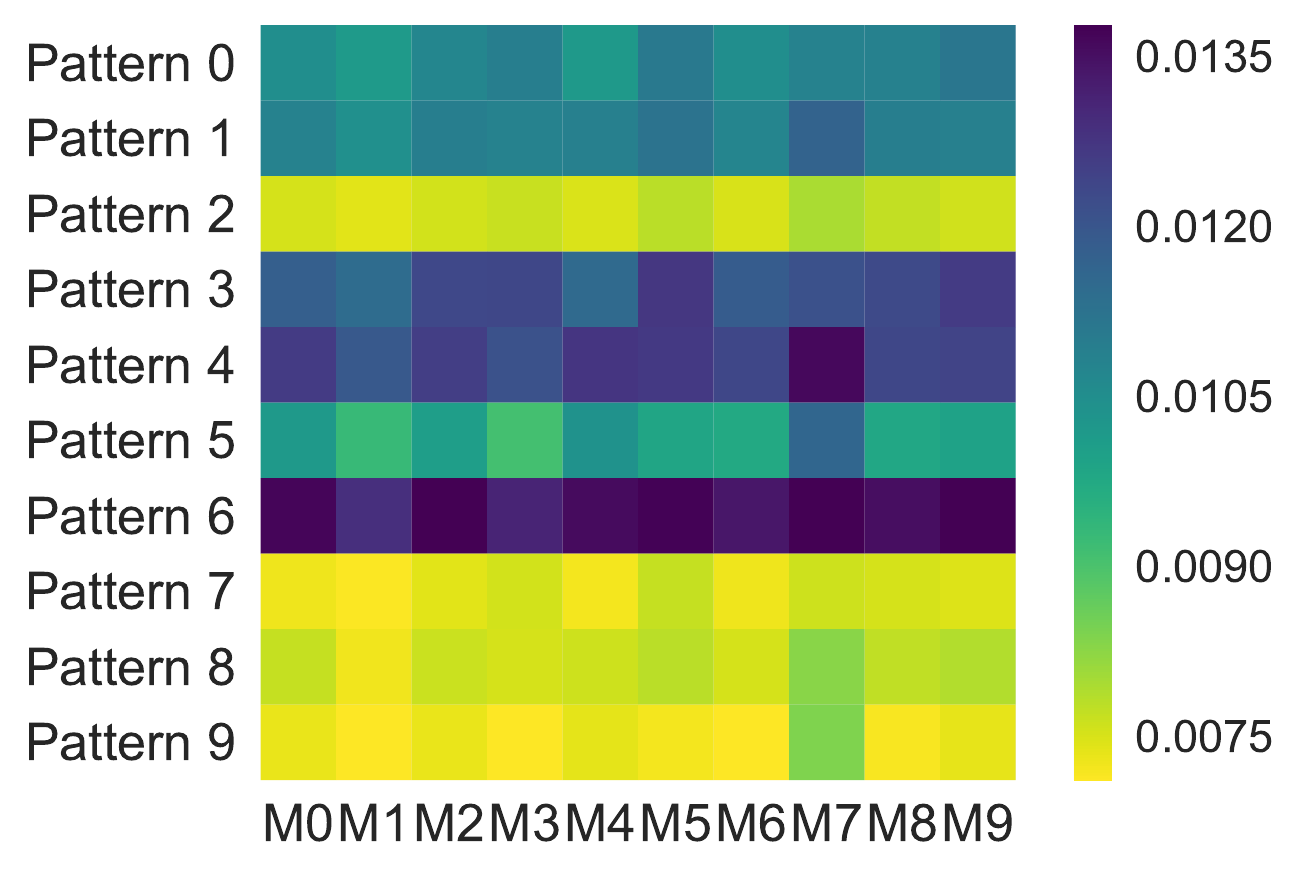}
	\caption{Attention weights over memory slices for patterns.}
	\label{fig:Attention}
\end{figure}

As introduced in Section \ref{Proposed_model}, the proposed architecture of TEMN incorporates topic information and attention mechanism in the memory network, which enables us to visualise the weighted importance of memory slices with respect to multiple patterns and thus enhance the interpretability of our model. We plot a heatmap of the weights in Figure \ref{fig:Attention}, where the colour scale represents the strength of the attention weights, and each row represents the average attention vector for each pattern. As can be observed, it is clear that each pattern has learned a certain type of selection rules across memory slices. Moreover, some patterns are particularly similar with each other than their counterparts. When we compared these results with the semantic meaning for each pattern as presented in Table \ref{tab:pattern}, we can find that similar patterns like 2, 7, 8, and 9 all have residential district as their top category. In contrast, pattern 6, which is related to Olympic venues, shows a distinctive attention vector, suggesting that the visiting features of this pattern can be extremely different from other types. 

\begin{table}[h]
	\setlength{\abovecaptionskip}{0 cm}
	\setlength{\belowcaptionskip}{0.35 cm}
	\renewcommand\arraystretch{1.25}
	\small
	\begin{tabular}{||c||c||}
		\hline 
		Pattern&Top Categories\\
		\hline
		\hline  
		$0$& Airport, Central business district, Hub station \\
		\hline  
		$1$ & Shopping mall, Square, Commercial street\\
		\hline  
		$2$ & University, Shopping mall, Residential district\\
		\hline  
		$3$ & Bar street, Stadium \\
		\hline  
		$4$ & Historic district, Shopping mall \\
		\hline  
		$5$ & High-tech zone, University, Shopping mall\\
		\hline  
		$6$ & Olympic Park\\
		\hline  
		$7$ & Residential district, SOHO\\
		\hline  
		$8$ & Hub station, Residential district\\
		\hline  
		$9$ & University, Residential district\\
		\hline 
	\end{tabular}
	\captionsetup{margin=0cm}
	\caption{Pattern and top categories.}
	\label{tab:pattern}
\end{table}

\subsubsection{Geographical Preferences of Patterns.}
\begin{figure}
	\begin{subfigure}[b]{0.48\columnwidth}
		\includegraphics[width=\linewidth]{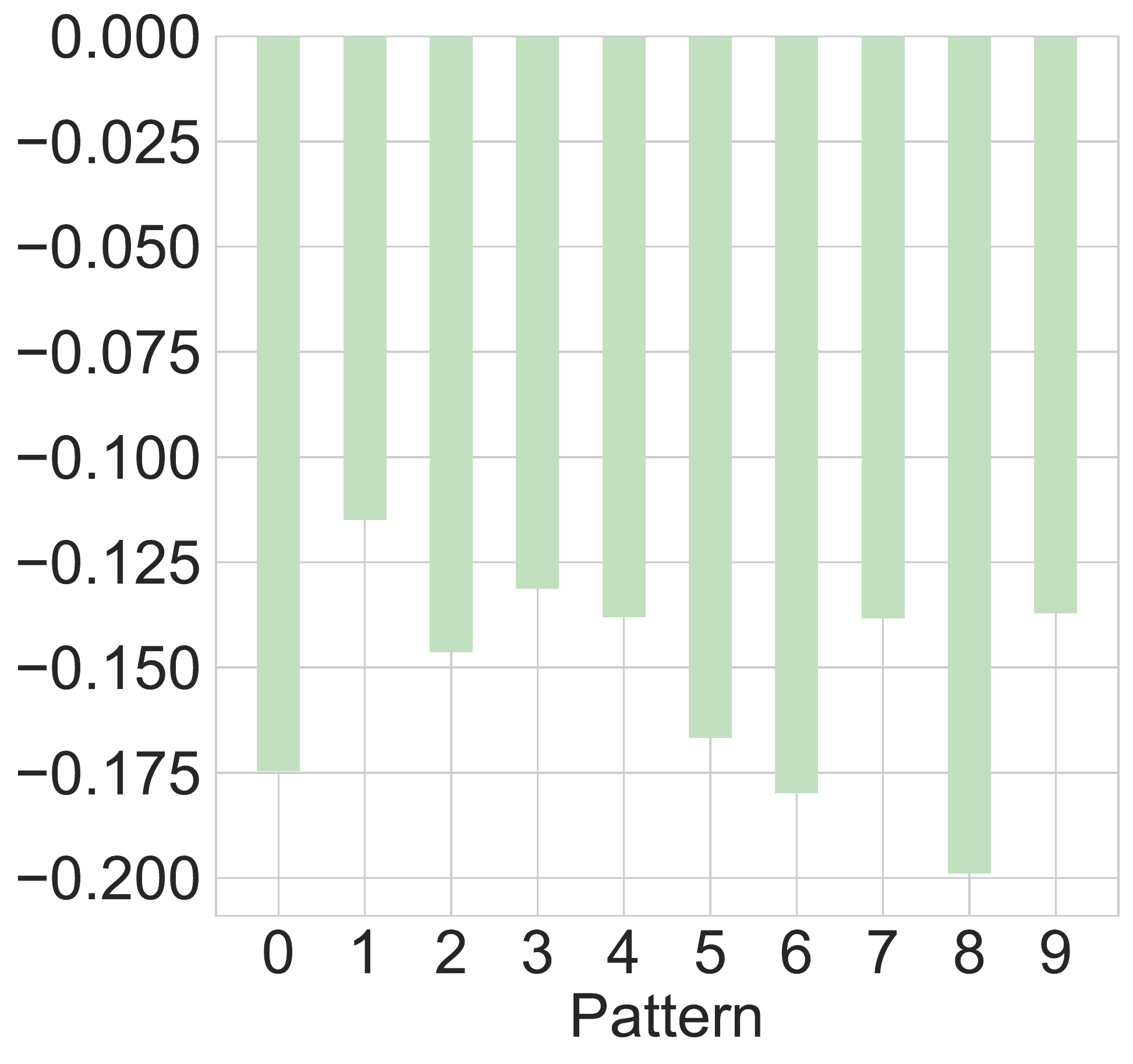}
		\caption{Geo-influence of POI $\rho_{v}$}
		\label{fig:1}
	\end{subfigure}
	\begin{subfigure}[b]{0.48\columnwidth}
		\includegraphics[width=\linewidth]{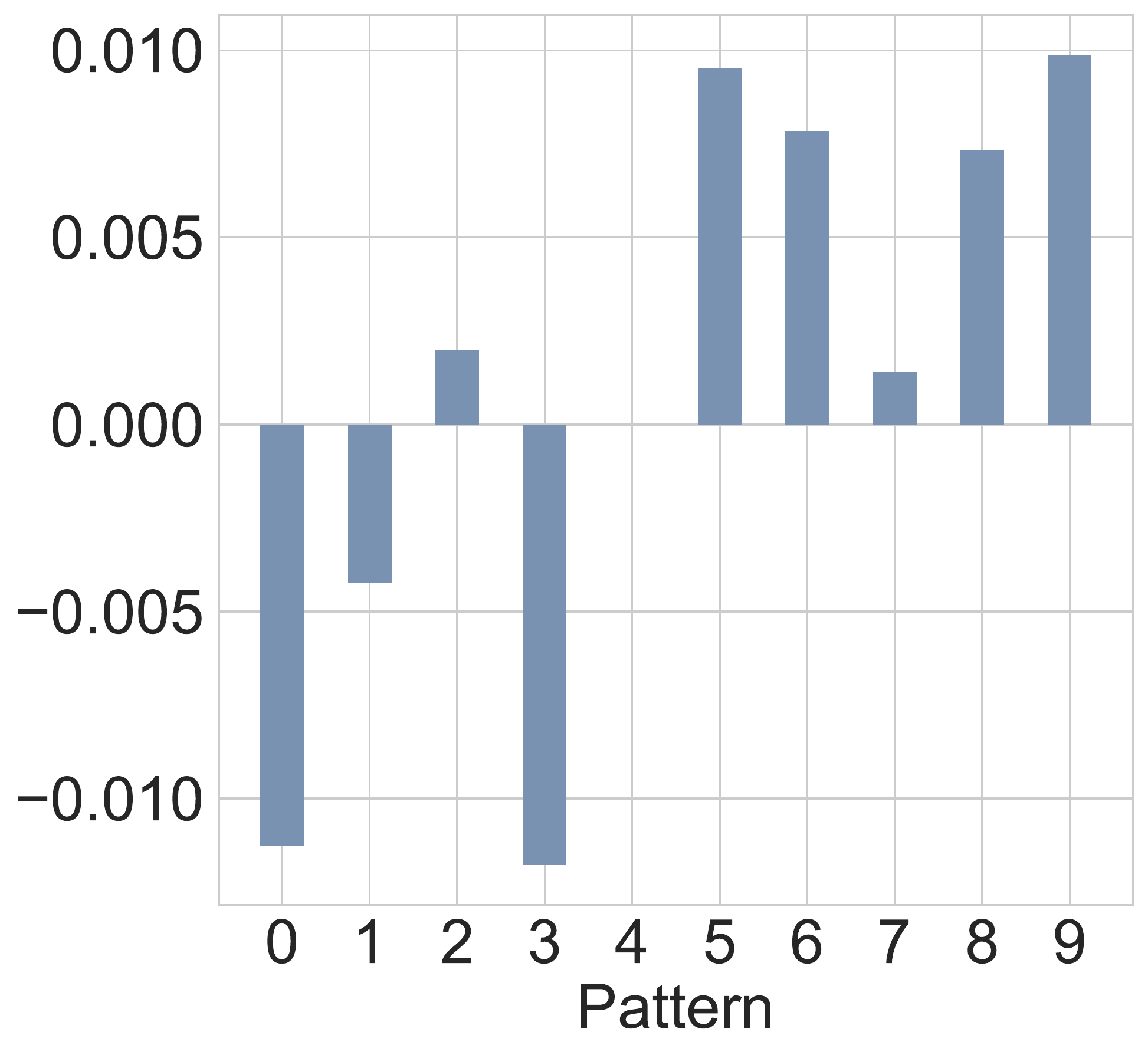}
		\caption{Geo-preference of user $\rho_{u}$}
		\label{fig:2}
	\end{subfigure}
    \caption{Average geographical influence of POIs and geographical preference of users for patterns.}
    \label{fig:Geo_para}
\end{figure}

As introduced in \ref{Geo_model}, the \textit{geographical influence} $\rho_{v}$ of a POI and \textit{geographical preference} $\rho_{u}$ of a user can be learned through our model. By calculating the average \textit{geographical influence} of POIs and \textit{geographical preference} of users for each pattern, we can gain insight into the geographical characteristics of different patterns. The calculated results are presented in Figure \ref{fig:Geo_para}, where a smaller number means the POI or user is less sensitive to distance. It can be seen that patterns of 0 and 8 give the smallest value of $\rho_{v}$, indicating that the visits to POIs belonging to these patterns (e.g. airport, central business district, hub station) are not influenced by distance too much; while for pattern 1, which is represented by POI categories more related to local services, it is rather sensitive to distance that people are unlikely to visit them from distant places. On the other hand, users from pattern 0 and 3 are relatively more willing to visit POIs far away than their counterparts by showing the smaller average $\rho_{u}$.

\section{Conclusion}
We introduce a novel end-to-end architecture, named Topic-Enhanced Memory Network (TEMN), for better POI recommendation with spatio-temporal contextual information considered. Different from existing deep learning frameworks for POI recommendation, TEMN is a unified hybrid model that leverages TLDA and external memory network with a neural attention mechanism to capture both global and fine-grained preferences of users. Comprehensive experiments under multiple configurations demonstrate the proposed architecture's significant improvements over competitive baselines. Qualitative visualisation of the attention weights and semantic investigation about patterns learned via TLDA module provide insight into the learning process and the recommendation results. 
In future work, we plan to extend TEMN to incorporate richer content information such as users' profiles and POIs' attributes.

\begin{acks}
We would like to thank Tencent for hosting Xiao Zhou and providing her with access to the datasets for this study. Our special appreciation also goes to Mr. Huanzhong Duan for providing necessary information regarding the datasets. The first author acknowledges the financial support co-funded by the China Scholarship Council and the Cambridge Trust.
\end{acks}

%
\bibliography{sample-base}
\appendix
\section{Appendix}

\begin{figure*}
	\vspace{0.8cm}
	\setlength{\abovecaptionskip}{0 cm}
	\setlength{\belowcaptionskip}{0.2 cm}
	\begin{subfigure}[b]{.244\linewidth}\label{fig:5a}
		\includegraphics[width=\linewidth]{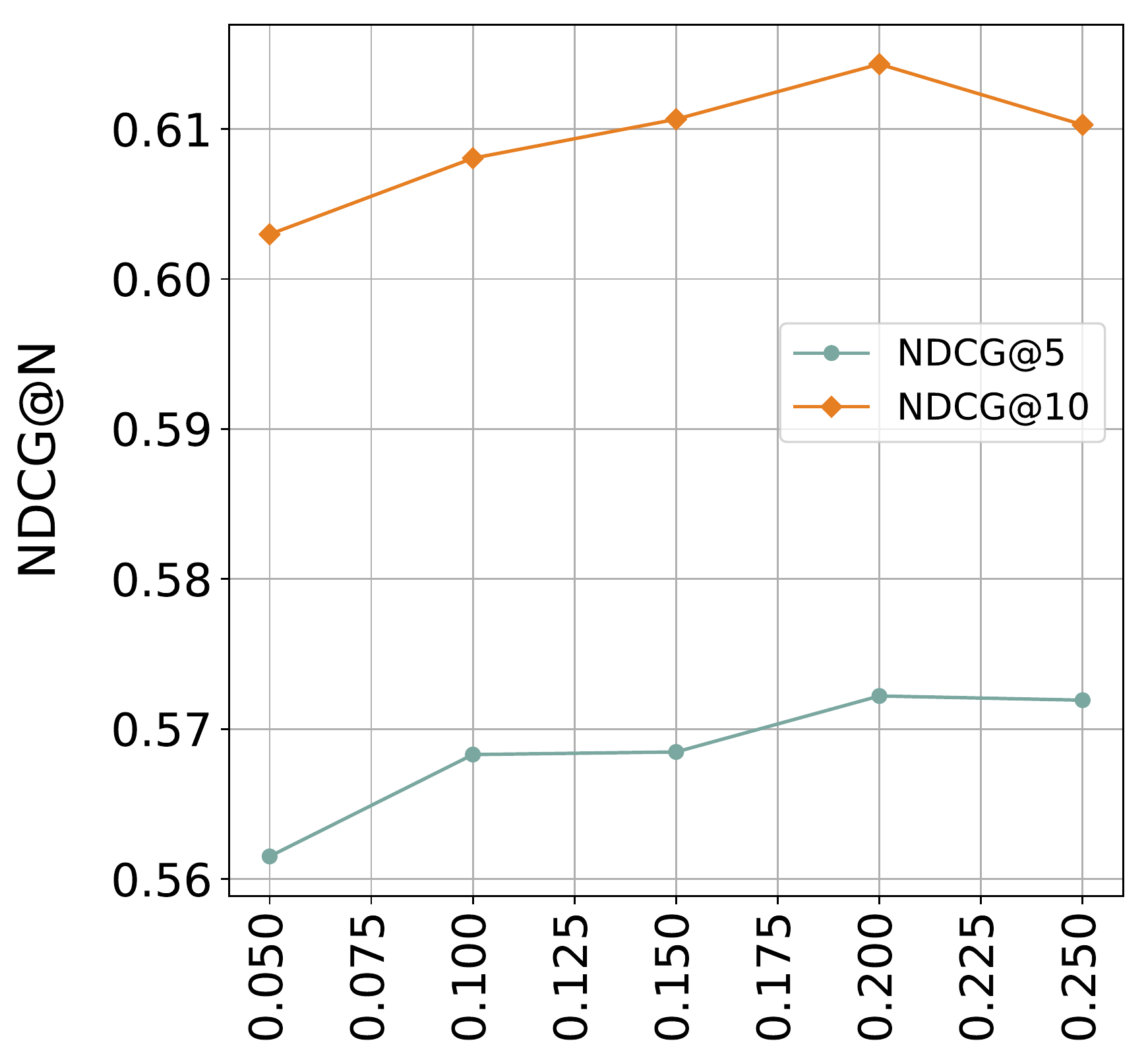}
		\caption{CPR--$\varsigma$}\label{fig:CPR_T_weight}
	\end{subfigure}
	\begin{subfigure}[b]{.247\linewidth}
		\includegraphics[width=\linewidth]{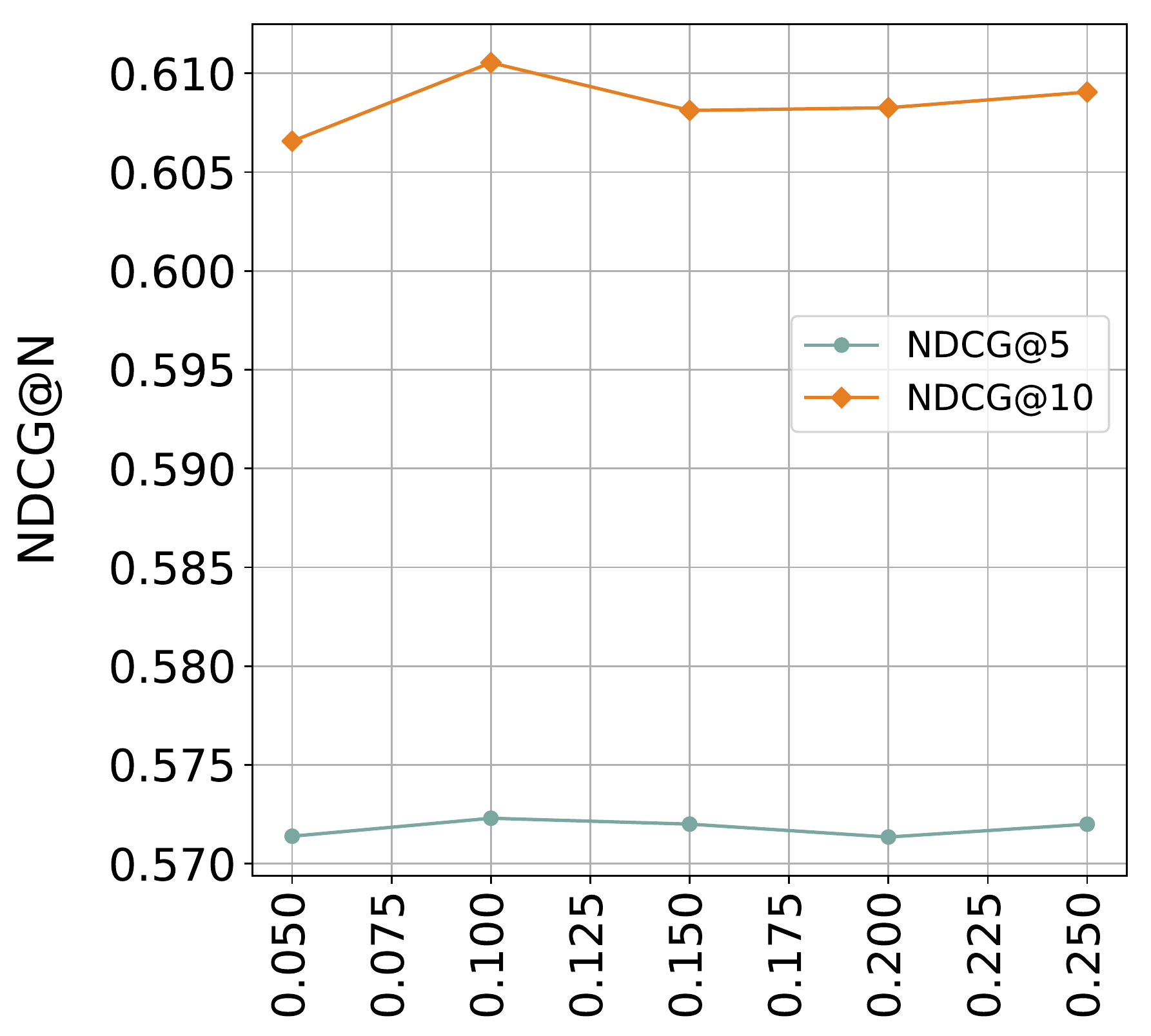}
		\caption{CPR--$\epsilon$}\label{fig:CPR_G_weight}
	\end{subfigure}
	\begin{subfigure}[b]{.24\linewidth}
		\includegraphics[width=\linewidth]{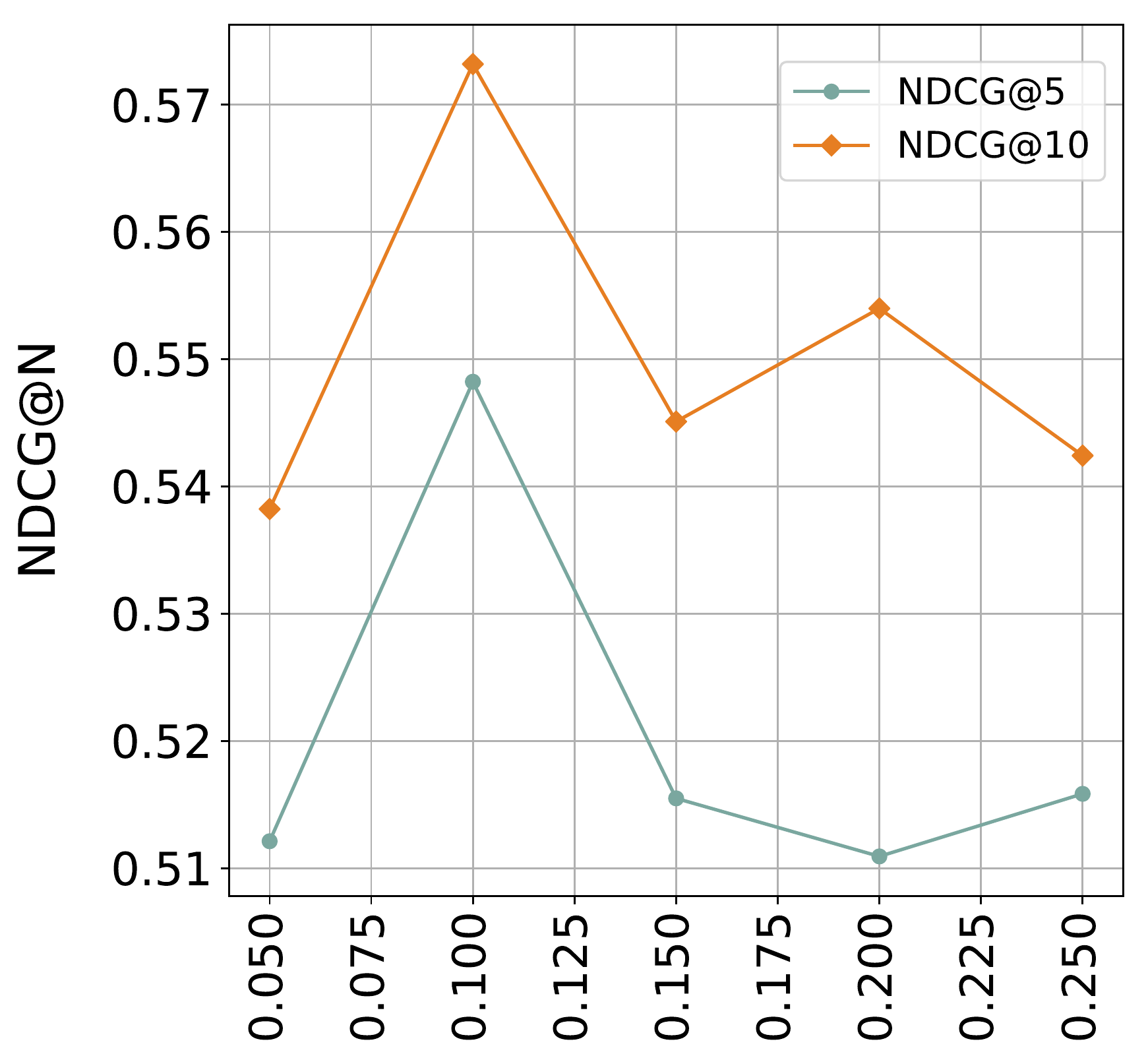}
		\caption{SPR--$\varsigma$}\label{fig:SPR_T_weight}
	\end{subfigure}
	\begin{subfigure}[b]{.254\linewidth}
		\includegraphics[width=\linewidth]{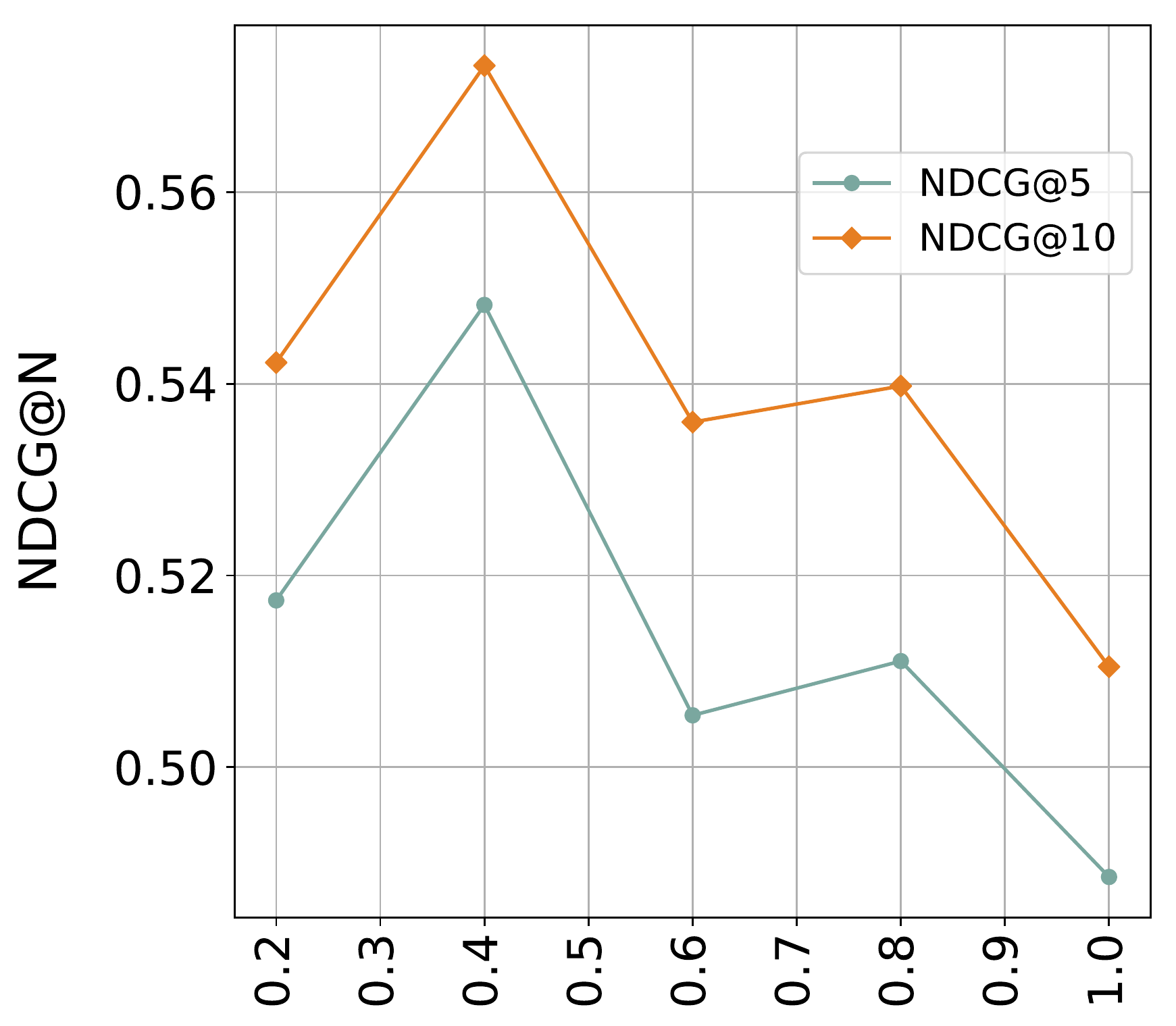}
		\caption{SPR--$\epsilon$}\label{fig:SPR_G_weight}
	\end{subfigure}
	\caption{Effect of weighting parameters in the loss function.}
	\label{fig:Loss_weights}
\end{figure*}

\begin{figure}
	\begin{subfigure}[b]{0.48\columnwidth}
		\includegraphics[width=\linewidth]{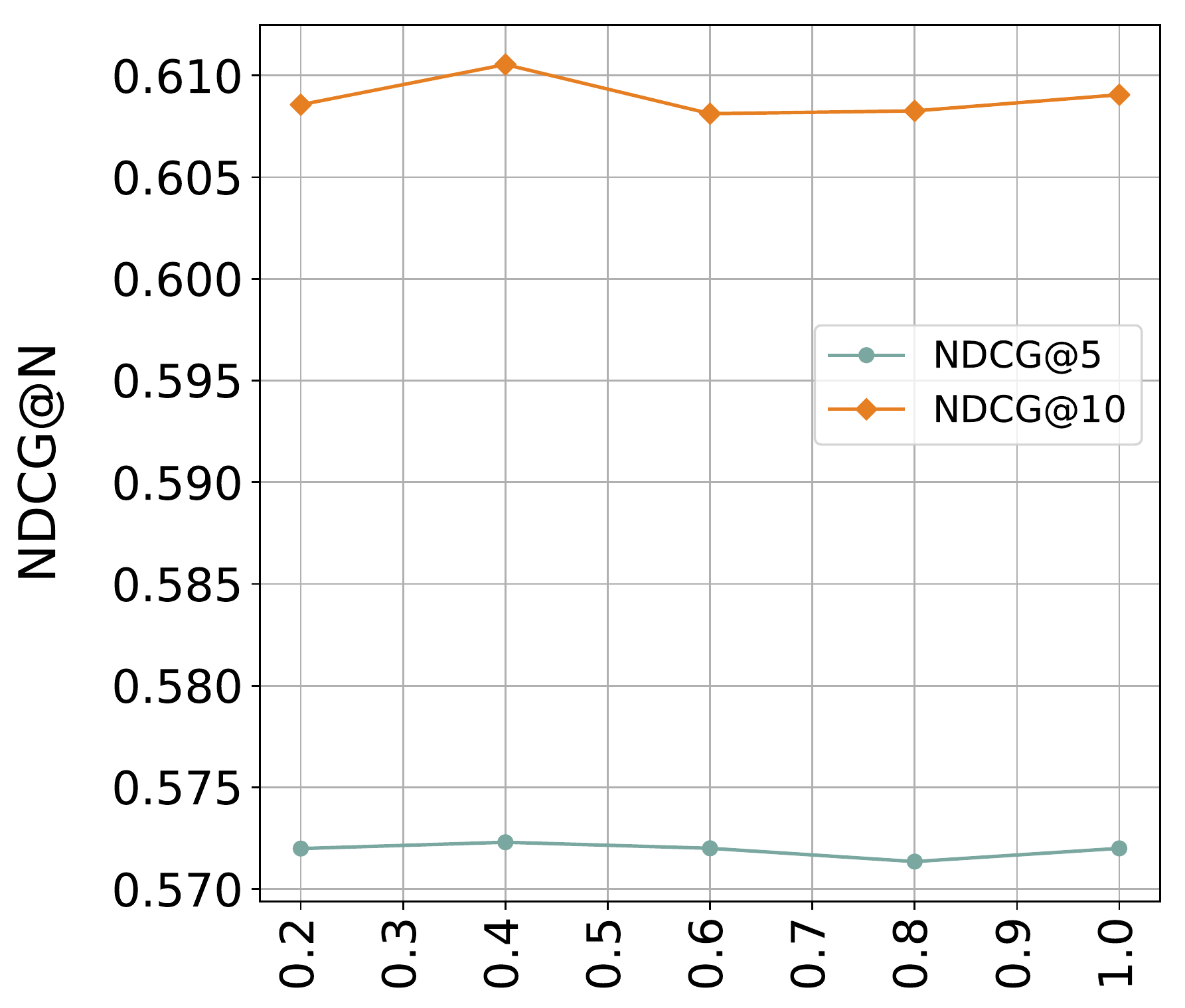}
		\caption{CPR--$\eta$}
		\label{fig:CPR}
	\end{subfigure}
	\begin{subfigure}[b]{0.48\columnwidth}
		\includegraphics[width=\linewidth]{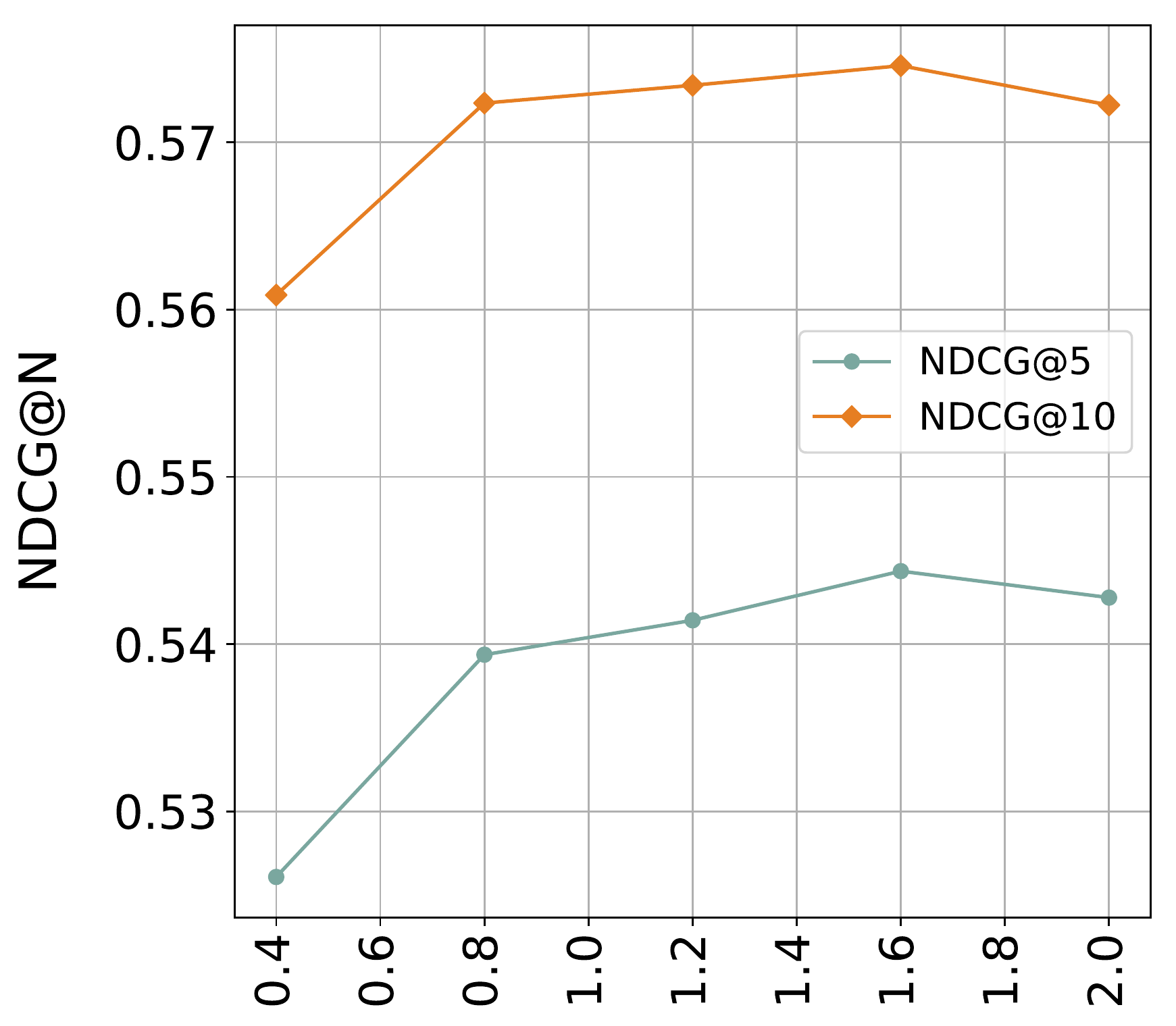}
		\caption{SPR--$\eta$}
		\label{fig:SPR}
	\end{subfigure}
	\caption{Effect of weighting parameter in score function.}
	\label{fig:Score_loss}
\end{figure}

 In this section, we provide detailed explanations of how we did our experiments for reproducibility. Our code is publicly available at https://github.com/XiaoZHOUCAM/TEMN.

\subsection{Dataset Filtering}
To ensure the quality of the data, we retain users who interacted with at least ten unique POIs. The filtered dataset is termed as \textit{WeChat(GPR)} and employed as the evaluation dataset for both general and context-aware recommendation. For sequential POI recommendation scenario, the dataset is further processed as that introduced in \textit{Sequential POI recommendation}, Section \ref{Problem_form}. More specifically, we set the threshold $\Delta T $ to one day (24 hours), and keep sequences with at least five successive check-ins. This newly created dataset containing the set of successive check-ins for each user is called \textit{WeChat(SPR)}.

\subsection{Evaluation Protocols}
In the experiments, the check-in data are transformed into binary implicit feedback as ground truth, indicating whether the user had visited the specific POI. For each dataset, we holdout the latest 15\% visiting history of each user to construct the test set, and split the remaining data into training (70\%) and validation (15\%) sets. The validation set is used to tune hyper-parameters and the final performance comparison is conducted on the test set. Closely following the setup that in \cite{he2017neural, rendle2009bpr}, we randomly sample 100 POIs that are not visited by the user and rank the test POI along with the 100 negative samples for evaluation.

\subsection{Parameter Tuning}
We implemented the TEMN model in TensorFlow\footnote{https://www.tensorflow.org/}. For our model, all hyperparameters are tuned according to the validation set based on the NDCG metric. We randomly initialise model parameters according to the uniform distribution and optimise the model by conducting stochastic gradient descent (SGD). The learning rate of SGD and the regularisation parameter are determined by grid search in the range of \{0.0001, 0.005, 0.001, 0.01\} and \{0.0001, 0.001, 0.01, 0.1\}, respectively. The dimensionality of user and item embeddings $d$ is tuned amongst \{20, 50, 75, 100\}. The number of memory slices in $M$ is tuned amongst $h$ = \{5, 10, 20, 50\}. And the number of patterns set in TLDA is tuned among \{3, 5, 10, 20\}. For MN module and geographical modeling that minimise the hinge loss, the margin is tuned amongst \{0.1, 0.2, 0.25, 0.5\}. 

\subsection{Effect of Weighting Parameters}
The weighting parameters utilised in the overall loss function (Equation \ref{all_loss}) and score function (Equation \ref{all_score}) are tuned by us to study the effect of incorporating TLDA module and geographical modeling for POI recommendation. Specifically, we study the performance of our models on NDCG@5 and NDCG@10 in context-aware POI recommendation (CPR) and sequential POI recommendation (SPR) scenarios by tuning the relavant weighting parameters as shown in Figure \ref{fig:Loss_weights} and Figure \ref{fig:Score_loss}.

It can be seen from Figure~\ref{fig:CPR_T_weight} and Figure~\ref{fig:SPR_T_weight}, when the weighting parameter of TLDA component $\varsigma$ is set to a small value of 0.05, models for CPR and SPR all give unfavourable results. When TLDA begins to play an increasingly important role, the performance is improved dramatically. The best results are achieved when $\varsigma\approx 0.2$ for CPR and $\varsigma\approx 0.1$ for SPR. However, when the weights of TLDA module continue to rise, the performance goes down. This result indicates that integrating TLDA module with memory network indeed helps to offer better POI recommendation, however, paying too much attention to user's intrinsic preference may weaken her neighbourhood-based characteristics. 

As for the influence of geographical modeling, an interesting observation is that in CPR scenario, the performance of models is stable when $\epsilon$ is tuned; while for SPR, the curves fluctuate substantially with the increasing of $\epsilon$. Moreover, it can be found that when the weighting parameters of geographical module are set to relatively larger values ($\epsilon\approx 0.4$ and $\eta\approx 1.6$), the performance of our models is most satisfying. These findings suggest that geographical effects have a greater influence on sequential POI recommendation, where shor-term preferences value more.

\subsection{Parameter settings}

Eventually, we find that the following hyperparameters work well: learning rate is set to 0.005, regularisation parameter is set to 0.0001, both of $\lambda^m$ and $\lambda^{\sigma}$ are set to 0.2. The dimension of user and item embeddings $d$ is set to 50, and the number of memory slots $h$ is set to 10. The number of patterns set in TLDA is 10. We set the weighting parameters $\varsigma$ = 0.2, $\epsilon$ = 0.1, and $\eta$=0.4 in general and context-aware recommendation; while for sequential POI recommendation, we set $\eta$ = 1.6 in the overall score function, $\varsigma$ and $\epsilon$ to 0.1 and 0.4, respectively.

\end{document}